\documentclass[prd,aps,eqsecnum,tightenlines,nofootinbib]{revtex4}
\input epsf
\usepackage{graphicx}

\newcommand{\beq}{\begin{equation}}
\newcommand{\beqa}{\begin{eqnarray}}
\newcommand{\eeq}{\end{equation}}
\newcommand{\eeqa}{\end{eqnarray}}

\newcommand{\lsim}{\lesssim}
\newcommand{\gsim}{\gtrsim}

\newcommand{\vect}[1]{\mbox{\boldmath${#1}$}}
\newcommand{\lmk}{\left(}
\newcommand{\rmk}{\right)}

\newcommand{\lkk}{\left[}
\newcommand{\rkk}{\right]}
\newcommand{\lla}{\left\langle}

\newcommand{\rra}{\right\rangle}

\newcommand{\vex}{{\vect x}}
\newcommand{\ver}{{\vect r}}

\newcommand{\vel}{\vect l}
\newcommand{\ven}{\vect n}

\begin{document}
\title{Annual modulation  of the Galactic binary confusion noise
bakground  and
LISA data analysis } 
\author{Naoki Seto
}
\affiliation{Theoretical Astrophysics, MC 130-33, California Institute of Technology, Pasadena,
CA 91125
}

\begin{abstract}
We study the anisotropies of the Galactic confusion noise background
 and its effects on LISA data analysis.  LISA has two  data
 streams of the gravitational waves signals relevant for low frequency
 regime. Due to the anisotropies of the background, the
 matrix for their confusion noises has off-diagonal components and
 depends strongly on the orientation of the detector plane. We find that
 the sky-averaged confusion noise level $\sqrt {S(f)}$ could change by a factor of 2
 in three  
 months, and would be minimum when the orbital position of LISA is
 either around the spring or autumn equinox.
\end{abstract}
\maketitle


\section{introduction}
The Laser Interferometer Space Antenna (LISA) is planned to  be launched
around 2011 and is expected to establish  a new window in the low
frequency 
gravitational 
wave astronomy from $0.1{\rm mHz}$ to $ 100$mHz
\cite{lisa,Cutler:2002me}. Its main 
astrophysical targets are  Galactic binaries and  cosmological
massive 
black holes (MBHs) that merge with other MBHs or capture
compact objects.  As  for the Galactic binaries, 
there would be  a lot of sources in the LISA band. For example we will  be
able to   
resolve several thousand  close white dwarf binaries at $f\gsim
3$mHz \cite{bcn1,bcn2}. At lower frequency regime $f\lsim  3$mHz they
are highly 
overlapped in the frequency bins, and it would be difficult to  resolve
them individually. As a result, they form a confusion noise background
whose magnitude 
could be 
larger than the detector noise at $0.1{\rm mHz}\lsim f \lsim 3$mHz
\cite{lisa}. 
The coalescence frequency  $f_c$ of a MBH system is given by its
redshifted 
total mass $M_z$ as $f_c\sim 2(M_z/2\times 10^6M_\odot)^{-1}$mHz. The
mass function of MBHs is highly uncertain at the lower end $\sim
10^5M_\odot$ \cite{Ferrarese:2000se}, and we might have to  search
treasurable signals  from 
cosmological  MBHs
in the Galactic confusion noise background.

The spatial distribution of the Galactic binaries would trace the
Galactic structure well,  and the confusion noise background would be
strongly anisotropic. The background itself can be regarded as a signal
that would provide some information on our Galaxy
\cite{gia,Cornish:2001hg,Ungarelli:2001xu}. But we should notice 
that 
there exists a more straightforward approach to probe the Galactic
structure using  thousands of 
resolved binaries whose three dimensional positions are estimated in some
error boxes \cite{gia}.  In this paper we will focus on the role of the
background as a noise, and study the effects of its anisotropies on the
signal  analysis of LISA. 
We do not pay attention to the normalization of the
background that has been studied by other papers using an 
isotropic approximation of the source distribution \cite{bcn1}.

This paper is organized as follows. 
In Section 2 we study the response of interferometers to the anisotropic
gravitational wave background with using the long wave approximation,
and formulate the noise matrix for the Galactic confusion background. In
Section 3 
we 
numerically evaluate the bakground noise as a function of the
orientation of the detector,  and estimate the annual modulation of the
noise level  for LISA. In Section 4 we discuss the anisotropies of the
signal to noise ratio (SNR) of sources at a fixed distance. Section 5 is
devoted to a 
brief summary of this paper. In the Appendix the confusion noise matrix
is analyzed  
in a  different manner from Section 2.

\section{formulation}

From  data streams of LISA,  we can make three meaningful  modes
$(A,E,T)$ such 
that the laser frequency noise can be reduced enough and their detector
noises do not correlate (see
\cite{prince} and references therein, also \cite{Cornish:2003tz}). At the
low frequency region ($f\ll 10$mHz: 
determined by the arm-length of LISA, $5\times 10^6$km)
relevant for the present analysis of the Galactic binary confusion noise,
 the $T$-mode has much worse sensitivity  than the rest.  Therefore we
only  discuss $A$ and $E$ modes  here.   These two data streams
$d_I~~(I=A,E)$ are made by the detector noise $n^D_{I}$ and the response
to  
gravitational  waves $h_{I}$ 
\beq
d_I=h_I+n^D_{I}.
\eeq
We define the detector noise spectrum  $S^D_{IJ}$ as follows \cite{thorne,cutfra}
\beq
\lla (n^D_{I}(f))^* n^D_{J}(f^{'}) \rra=\frac12 \delta(f-f^{'})S^D_{IJ}(f)\label{sd}.
\eeq
The frequency dependence is irrelevant for our analysis, and, hereafter,  we omit its
explicit dependence for  notational simplicity. We have the following
relation for the detector noise spectrum matrix
\beq
S^D_{AA}=S^D_{EE}, ~~~ S^D_{AE}=0.
\eeq

The three spacecrafts of LISA form a nearly equilateral triangle, but
due to  the symmetric combinations of data,
the two responses  $h_{I}$ $(I=A, E)$ to  the low frequency gravitational  waves can be essentially
regarded as that of two  $90^\circ$-interferometers rotated by $45^\circ$
from each other, as shown in figure 1. We take the reference coordinate
system $ (X_D,Y_D,Z_D)$. The $Z_D$-axis is
normal to the 
detector 
plane.  
When we define two units vectors $\vel$ and $\ven$,
the response of the $A$ mode is proportional to $\vel \cdot H \cdot
\vel -\ven \cdot H \cdot \ven$ and the $E$ mode to $2\ven \cdot H \cdot
\vel  $ with some tensor $H$ made from gravitational waves.
 It is important to  note that there is one degree of freedom to
the data combinations $(A,E)$ \cite{Nayak:2003na}. We can  make the following linear
combinations $(d_A(\phi_f), d_E(\phi_f))$ made from the original modes
$(d_A,d_E)$ 
with a two dimensional  rotation matrix $R(\phi)$  as
\beq
\left( 
           \begin{array}{@{\,}c@{\,}}
           d_{A}(\phi_f)    \\
             d_{E}(\phi_f)   \\ 
           \end{array} \right)
 =R(2\phi_f)
\left( 
           \begin{array}{@{\,}c@{\,}}
           d_{A}    \\
             d_{E}   \\ 
           \end{array} \right). \label{trans}
\eeq
These new modes $(A(\phi_f), E(\phi_f))$ are equivalent to  the responses
of two $90^\circ$-interferometers that are
obtained by 
rotation of the original  modes  $(A,E)$ with angle $\phi_f$ around the
$Z_D$-axis.  This reflects  the spin-2  nature of  a gravitational wave.
We can easily confirm the following relations for the detector noise
spectrum 
\beq
S^D_{AA}(\phi_f)=S^D_{EE}(\phi_f)=S^D_{AA}=S^D_{EE},~~~S^D_{AE}(\phi_f)=0.
\eeq
Due to the simple  relation (\ref{trans}) this freedom of the effective
rotation 
does not affect our final  results as we see later. Therefore  we simply
use the original  modes $(A,E)$ for most of our analysis.

Now we move to the binary confusion noise  that is made by a large number
of unresolved 
Galactic binaries. At the low frequency regime the response of the
mode
$A$ to  a monochromatic binary is given by two polarization modes
$(+,\times)$ of gravitational waves  \cite{thorne}
\beq
h_A=a^+ F_A^+   +a^\times F_A^\times. \label{one}
\eeq
With the quadrupole formula the amplitudes $a^+$ and $a^\times$ are
expressed as  
\beq
a^+=K(1+\cos^2\theta_i), ~~~~a^\times=2K \cos\theta_i,
\eeq
with  the inclination
angle $\theta_i$, and the coefficient $K$ is given in terms of the distance to  the binary $r$, the chirp mass $m_c$, and the
gravitational  wave frequency $f$ as
\beq
K=2 \frac{G^{5/3} m_c^{5/3}}{r c^4} (\pi f)^{2/3}.
\eeq 
Two functions $ F_A^+$ and  $ F_A^\times$ are determined by the sky
position $(\theta,\phi)$ and the polarization angle $\psi$ of the binary
in the $(X_D,Y_D,Z_D)$ coordinate as
\beqa
F_A^+&=&\frac12 (1+\cos^2\theta) \cos2\phi \cos2\psi-\cos\theta \sin2\phi
\sin2\psi,\label{f1}\\
F_A^\times&=&\frac12 (1+\cos^2\theta) \cos2\phi \sin2\psi+\cos\theta \sin2\phi \cos2\psi.\label{f2}
\eeqa 
The response $h_E$ to  the $E$ mode is given by a similar expression as
eq.(\ref{one}). We replace the azimuthal angle $\phi$ to $\phi-\pi/4$ in
eqs.(\ref{f1}) and 
(\ref{f2}). 
As in eq.(\ref{sd}) defined for the detector noise,
we denote  the Galactic binary noise spectrum matrix $S^B_{IJ}$ by the
responses $(h^B_{A}, h^B_{E})$ of the modes $(A,E)$ to  all  the
unresolved Galactic 
binaries. 
With using eq.(\ref{one}) we sum up all the unresolved binary  within 
Galaxy.
For simplicity we assume that the binary mass function does not depend
on the frequency 
$f$ or the position in the Galaxy. This would be a fairly reasonable
assumption. Considering
the nature of the confusion noise we average over the inclination and
polarization  angles $\theta_i$, $ \psi$, the spatial
position $\ver$,  and obtain the following expression 
\beq
S^B_{IJ}=P \int d\ver \rho (\ver) r^{-2}
F_{IJ}(\theta,\phi), \label{sb}
\eeq
where $P$ is a normalization factor that is not important for our
analysis. The function $\rho(\ver)$ is the density profile of the Galactic
binaries, and the functions $F_{IJ}$ are defined as
\beqa
F_{AA}&=&\frac12\lkk \lmk \frac{1+\cos^2\theta }2 \cos 2\phi \rmk^2
+\lmk\cos\theta \sin 2\phi \rmk^2   \rkk, \\
F_{EE}&=&\frac12\lkk \lmk \frac{1+\cos^2\theta }2 \sin 2\phi \rmk^2
+\lmk\cos\theta \cos 2\phi \rmk^2   \rkk, \\
F_{AE}&=&F_{EA}=\frac12\lkk\lmk \frac{1+\cos^2\theta}2
\rmk^2   -\cos^2\theta
  \rkk \cos2\phi \sin2\phi.
\eeqa
Using a spherical coordinate we can express  eq.(\ref{sb}) as \cite{gia}
\beq
S^B_{IJ}=\int d\Omega B(\theta,\phi)F_{IJ}(\theta,\phi),
\eeq
where the anisotropy of the gravitational  wave background
$B(\theta,\phi)$ is defined as 
\beq
B(\theta,\phi)\equiv P \int dr  \rho(r, \theta,\phi).\label{gwaniso}
\eeq 
If the source distribution $\rho(\vex)$ for gravitational waves is same
as that of the light and the interstellar absorption is negligible, the
function $B(\theta,\phi)$ is also  the luminosity distribution of the
Milky Way on the sky.

In reality there would be some fluctuations around the above expression
(\ref{sb}) 
due to the finite number of the binaries. But this effect is
beyond the scope of this paper.  Some binaries would be close enough to have
significant SNR above the confusion noise level, and could be removed
from the 
background. But a rough estimation indicates that the typical distance to
such  binaries is smaller than $\sim 100$pc at the low frequency
region  $f<1$mHz, and their subtraction
would change our results only slightly \cite{Seto:2002dz}. 

For an isotropic binary distribution   we can replace the function
$F_{IJ}$ by
\beq
F_{AA}=F_{EE}=\frac15,~~~F_{AE}=0,
\eeq
and the noise matrix $S^B_{IJ}$ is diagonal (proportional to the unit
matrix). For reference we define the noise amplitude $S^B_{iso}$ that is
obtained  with
using $F_{IJ}=1/5$ in eq.(\ref{sb}).
This monopole spectrum $S^B_{iso}$ is what has been used in most of previous
studies.  Differences between $S^B_{IJ}$ and $S^B_{iso}$ due to the
proper angular  dependence of 
$F_{IJ}$ are the main issue of this paper.

We can calculate the noise matrix $S^B_{IJ}(\phi_f)$ for the new data
combination 
$(A(\phi_f)$, $E(\phi_f))$ generated by  the simple relation
(\ref{trans}). 
 It is given by the original one $S^B_{IJ}$ as
\beq
S^B_{IJ}(\phi_f)=R(2\phi_f) S^B_{IJ} R(-2\phi_f).
\eeq
Let us  consider an analysis for some strong gravitational wave signal
$(h_{A,X}, h_{E,X})$ from a source $X$, 
{\it e.g.} the ring 
down waveform from  a merged MBH binary.
When the detector noise $S^D_{IJ}$ is much smaller than the Galactic
confusion noise $S^B_{IJ}$,  the SNR is given as
\beq
SNR^2\propto (h_A, h_E)^* (S^B )^{-1}
\left( 
           \begin{array}{@{\,}c@{\,}}
           h_{A}   \\
             h_{E}  \\ 
           \end{array} \right).\label{trans2}
\eeq
From eqs.(\ref{trans}) and (\ref{trans2}) we can easily confirm that the freedom of the
rotation angle $\phi_f$ is completely killed in the above relation for
SNR. 
This is the reason we can forget it from the beginning.
But it is instructive to  take a rotation angle $\phi_f=\phi_0$ so  that
the noise 
matrix $S^B_{IJ}(\phi_0)$ is diagonalized as 
\beq
\left( 
           \begin{array}{@{\,}cc@{\,}}
           \lambda_1 & 0  \\
            0 & \lambda_2 \\ 
           \end{array} \right).
\eeq
Then we have the following simple relation
\beq
SNR^2\propto \frac{|h_A(\phi_0)|^2}{\lambda_1}+\frac{|h_E(\phi_0)|^2}{\lambda_2}.
\eeq
Next we evaluate the effective noise level for the two modes by averaging
out  
 the direction and orientation the source $X$.
Apparently the  averages for $|h_{AX}(\phi_f)|^2$ and $
|h_{EX}(\phi_f)|^2$ do
not depend on the choice of the angle $\phi_f$, and we have $\lla
|h_{AX}(\phi_0)|^2 
\rra=\lla |h_{EX}(\phi_0)|^2 \rra$.  
Thus it is reasonable to  define the effective noise level $S^B_{eff}$
by the 
following equation 
\beq
S^B_{eff}=2\lmk \frac1\lambda_1+\frac1\lambda_2\rmk^{-1}=2\frac{\det S^B}{{\rm tr}S^B}.\label{dia}
\eeq
The coefficient 2 comes from the number of the modes.
The inverse of this quantity is proportional to  the averaged $SNR^2$ for
sources at a fixed distance. For  ground-base detectors, such as,  LIGO
II,  
the effective distance $d_e$ to a source that can be detected above a
given SNR threshold is often used to characterize the noise level.  When
the 
cosmological  effects for the event rate is small ({\it
e.g.} $d_e\sim 
300$Mpc), this is a intuitive measure for a detector.  But for the
cosmological  sources the situation is not so simple. Therefore we
do not pursuit such a kind of measure,
but use 
the effective noise 
$S^B_{eff}$. This approach greatly simplifies the treatment of the
polarization angle.

Finally we make a brief mathematical comment on the
effective noise 
level $S^B_{eff}$.
A  rigid analysis is given in the Appendix.
The functions $F_{IJ} (\theta,\phi)$ are written by spherical harmonic
functions 
 $Y_{lm} (\theta,\phi)$ with $l=0$, 2 and 4 \cite{Cornish:2001hg,Ungarelli:2001xu}. This means that the angular
dependence of the Galactic confusion noise felt by the detectors $(A,E)$
is 
limited to  these three modes at low frequency region. 
Therefore the effective noise for arbitrary detector plane
is determined by     $1+5+9=15$ real
parameters that 
characterize anisotropies of the binary noise (total of 5+9) and the
monopole 
mode (total of 1). The monopole mode is nothing but the isotropic
component $S^B_{iso}$.

\section{annual modulation of the effective noise}

Due to the annual rotation of the detector plane of LISA, the effective
noise $S^B_{eff}$ changes with time. To  denote the orbital phase of
LISA we use the 
time $T$ measured from the autumn equinox point in units of year. The
actual  date for $T=0$ corresponds to $\sim20$ days after the autumn
equinox day,  as LISA trails $20^\circ$ behind the Earth.
With eq.(\ref{sb}) we can calculate the annual modulation of the noise
$S^B_{eff}$ at given moment $T$. The detector plane  inclines to  the
ecliptic plane by $60^\circ$. We assume that the vertex of the cone made
by the envelope of the detector planes exists at south as shown in
figure 2.  Hereafter we implicitly assume this configuration.
When we take it to  the opposite direction, the time
dependence of the noise $S^B_{eff}$ changes by $T=0.5$yr from our results.
In figure 2 we also define the ecliptic coordinate that is very useful
for describing the  motion of LISA. In this coordinate the directions of the Galactic center, the
north and south Galactic poles are
$(\theta,\phi)=(-95.5^\circ,-93.2^\circ)$, $(60.4^\circ,180.0^\circ)$ and
$(119.6^\circ,0.0^\circ) $ \cite{GA}.

\subsection{Galactic model}
First we calculate the all sky map  of the gravitational wave luminosity $B(\theta,\phi)$.
We mainly use the Galactic stellar distribution model given in \cite{Binney:1996sv}.
This model (hereafter BGS model) contains both the triaxial bulge and disk components obtained
by fitting the near infrared COBE/DIRBE surface brightness map. In
addition to 
these two components we have also  studied contribution of the  halo
stars. But its effect is almost negligible for reasonable model
parameters, and we do not discuss the halo component in this paper (see
also \cite{gia}). The
explicit form  
of the BGS model is given as
\beq
\rho(\vex)=C (f_b(\vex)+f_d(\vex)),
\eeq
where $C$ is a normalization constant and
\beqa
f_b&=&f_0 \frac{e^{-a^2/a_m^2}}{(1+a/a_0)^{1.8}},\\
f_d&=&\lmk\frac{e^{-|z|/z_0}}{z_0}+\alpha \frac{e^{-|z_|/z_1}}{z_1}
\rmk R_s e^{-R/R_s},\\
a&\equiv& \lmk x^2+\frac{y^2}{\eta^2}+\frac{z^2}{\zeta^2}  \rmk, \\
R&\equiv& (x^2+y^2)^{1/2}.
\eeqa
Here we use a Galactic coordinate. Its $z$-axis is oriented to the north
Galactic pole, and the direction 
of the $x$-axis (the long axis of the triaxial bulge) is $20^\circ$
different from the Sun-center line. In this
model the
 solar system exists at  $R=8.0$kpc and $z=14$pc. The parameters are
given as 
$f_0=624, a_m=1.9$kpc, $a_0=100$pc, $R_s=2.5$kpc, $z_0=210$pc,
$z_1=42$pc, $\alpha=0.27$, $\eta=0.5$ and $\zeta=0.6$.   We performed numerical integration of
eq.(\ref{gwaniso}) and  obtain the map $B(\theta,\phi)$ as shown
in figure 3 where  the  ecliptic coordinate is used. We can
observe the bulge structure around the Galactic center.  Note that, for the
configuration of 
figure 2, the $Z_D$-axis of  LISA passes pretty close to  the north Galactic pole around the
autumn equinox. 

\subsection{noise map}

With the map $B(\theta,\phi)$ of the gravitational  wave anisotropies
the effective noise map $S^B_{IJ}$ is obtained by convolution of the
response functions $F_{IJ}$. 
In figure 3 
the noise map $S^B_{eff}$ is given as a function of the orientation of
the $Z_D$-axis 
(with angles $(\theta,\phi)$ in the ecliptic coordinate) of the
detector. Two  orientations $(\theta,\phi)$ 
and 
 $(\pi-\theta, \phi+\pi)$ give a same value due to the symmetry of the
response functions $F_{IJ}$.
The $Z_D$-axis of LISA moves along the dot-dashed line.
  In the map the maximum value of
${S}^B_{eff}/S^B_{iso}$ is 1.69 around the direction of the Galactic
center, and minimum 
value is $0.38$ around  the Galactic poles. 
As mentioned before the $Z_D$-axis of LISA passes pretty close
to the poles at the autumn equinox. 
This means that the effective noise
at the autumn equinox is almost the smallest value in the map.
We can also see
that the effective noise is higher in summer than winter. The separation
between the $Z_D$-axis and the Galactic center on the sky
is smaller for  the former. The map traces the
stellar distribution along the Galactic disk (Milky Way on the sky), but
is not  as significant as figure 3.

For the BGS model we also analyzed the deformation of the noise
matrix $S^B_{IJ}$  from a following  simple shape that is proportional  to the
unit matrix 
\beq
\left( 
           \begin{array}{@{\,}cc@{\,}}
           \lambda & 0  \\
            0 & \lambda \\ 
           \end{array} \right).
\eeq
 This deviation is caused
by the hexadecupole ($l=4$) mode of the
background  $B(\theta,\phi)$ (see Appendix).
We numerically calculate the ratio
$|\lambda_1-\lambda_2|/(\lambda_1+\lambda_2)$ for all  the sky
orientation of the $Z_D$-axis, and  found that its maximum value is
0.152.  Therefore it is not a
bad approximation to  put $\lambda_1=\lambda_2=\lambda$,  namely
$S^B_{AA}=S^B_{EE}=\lambda$ and $S^B_{AE}=S^B_{EA}=0$. In this
approximation the noise matrix is characterized by a single quantity
$\lambda(\theta,\phi)$ and Figure 3 can be roughly regarded as its map.

We have adopted the BGS model for the spatial distribution of the
Galactic binaries. But the radial scale length  of the disk $R_s$ has
some uncertainties \cite{Sackett:1997wf}.  Furthermore the
distribution of the gravitational wave sources might be different from
that of the near infrared light. Here we study how the effective noise
changes with 
the model parameter.
We use the following
exponential  disk model for  the Galactic density profile $\rho(\ver)$ \cite{Sackett:1997wf} 
\beq
\rho(R,z)=n_0 \exp[-(R/R_s)] {\rm sech}^2[(z/z_s)], \label{disk}
\eeq
with the Galactic cylindrical coordinate $(R,z)$ and a normalization
constant 
$n_0$. The radial scale length $R_s$ has typically $R_s= 1.7\sim
3$kpc. In this model we fix the disk scale height at $z_s=200$pc, and put the solar
system at $R=8.5$kpc and $z=30$pc.

In figure 4 we show our numerical results for the normalized noise level
$S^B_{eff}/S^B_{iso}$ as a function of the time $T$.  We choose the scale  length $R_s$ at $R_s=1.0,1.7$
and 2.5kpc. The results for $R_s=2.5$kpc and the BGS model are almost the
same.   
  As in figure 2 the  Galactic center is nearly on the
ecliptic plane and also nearly normal to  the two equinox
directions. Let us first consider a simple situation that (i) 
 the density profile is a delta function $\delta^3(\ver-\ver_{GC})$
around the 
Galactic center $\ver_{GC}$, and (ii)  the center is on the detector plane
with  angles $(\theta,\phi)=(\pi/2,\phi_{GC})$ in the ($X_D,Y_D,Z_D$)
coordinate 
system in figure 1.   Then the mode $A(\phi_{GC})$ is completely free
from the Galactic binary 
noise and one of the eigen values $\lambda_1$ vanishes. Thus the effective
noise becomes  $S^B_{eff}=0$ from eq.(\ref{dia}).
In reality the density profile $\rho(\ver)$ has a three dimensional
structure and the eigen values $\lambda_i$ take  finite values.  But the
density profile 
are highly concentrated around the center $\ver_{GC}$, and 
the minimum values of $S^B_{eff}/S^B_{iso}$ becomes smaller
as we decrease the scale length $R_s$ as shown in figure 5. The
deformation 
$|\lambda_1-\lambda_2|/(\lambda_1+\lambda_2)$ is larger for a  smaller
length $R_s$.


As the effective noise level $S^B_{eff}$ changes with time $T$, it would
be 
useful to  define a time average of the noise.
The most natural definition from   the standpoint of the signal analysis  would be the
following one 
\beq
\bar{S}^{B}_{eff}\equiv \lmk\frac1{\rm 1 yr} \int_0^1\frac{dT}{S^B_{eff}(T)}  \rmk^{-1}. 
\eeq
This quantity is inversely proportional to the all sky average of
$SNR^2$ for sources at a fixed distance as the effective noise $S^B_{eff}$.
We have
$\bar{S}^B_{eff}/S^B_{iso}=1.54$ and $1.15$ 
 for $R_s=1.0$ and $2.5$kpc respectively. Therefore the traditional
estimation based 
on the monopole mode gives a fairly good result for the time averaged
noise.

\section{snr as a function of the sky position}
So far we have discussed how the binary noise changes with the
orientation of the detector. 
In this section we first analyze the dependence of SNR on the sky
position of a monochromatic
source with a fixed distance and 1yr integration time.  We denote the source direction
$(\theta_s,\phi_s)$ in the ecliptic coordinate. With the rotation of
LISA,  
 the noise matrix $S^B_{IJ}(t)$ and the response function
$F^{+,\times}_{A,E}$  both change with time. At a given moment, the
direction 
of the binary $(\theta_{D},\phi_{D})$ in the rotating detector
coordinate $(X_D,Y_D,Z_D)$ can be formally  expressed as
($\theta_{D}(\theta_s,\phi_s,t)$, $\phi_{D}(\theta_s,\phi_s,t)$). 
Then the quantity $SNR^2$ with averaging over the source orientation
(polarization 
and inclination) is proportional to the following time integral
\beq
G(\theta_s,\phi_s)=\int_{\rm 1yr} \sum_{IJ}
F_{IJ}(\theta_{D}(\theta_s,\phi_s,t), \phi_{D}(\theta_s,\phi_s,t))
(S^B_{IJ} (t))^{-1} dt \label{snint}
\eeq
 Actually the polarization angle $\psi$ in eqs.(\ref{f1}) and (\ref{f2}) also changes with time. But its
average commutes with the time integration.

To deal with the complicated time dependence of the response function
$F_{IJ}$ and 
the noise matrix $S^B_{IJ}(t)$ we use  the following three prescriptions with

(1)  the monopole noise model $S^B_{IJ}(t)=S^B_{iso}\delta_{IJ}$ and
the angular averaged response functions $F_{IJ}=\delta_{IJ}/5$,

(2)  the monopole noise model $S^B_{IJ}(t)=S^B_{iso}\delta_{IJ}$ but
including the time dependence of the response functions $F_{IJ}$,

(3) including the time dependence for both the noise matrix $S^B_{IJ}(t)$ and the response
functions  $F_{IJ}$.

We denote the integral $G$ with above prescriptions as $G_1$, $G_2$ and
$G_3$ respectively.  The method (2) has been sometimes used for study of
the signal analysis that would be valid at higher frequency regime where
the 
detector noise dominant \cite{Cutler:1998ta}. As  the configuration of
LISA is symmetric around the 
$Z_E$-axis of the ecliptic coordinate (figure 2) and the freedom of the
$\phi_f$-rotation is irrelevant here, the result $G_2(\theta_s, \phi_s)$
should depend only on the zenith angle $\theta_s$.  We can
derive a 
simple analytic expression for the ratio $G_2/G_1$ as follows
\beq
\frac{G_2(\theta_s)}{G_1}={\frac{15605-2300\cos 2\theta_s -185\cos 4 \theta_s  }{16384}}.
\eeq 
The function ${G_2(\theta_s)}/{G_1}$ becomes maximum  1.08 at
$\theta_s=\pi/2$ and minimum  0.80 at $\theta_s=0$ and $\pi$.  Due to
the  angular average of the response function by the
annual rotation of LISA,  the simple prescription (1) gives fairly
good results for the isotropic noise.
We can
easily confirm that the angular average of the ratio  ${G_2(\theta_s)}/{G_1}$
becomes 
unity.

Next we calculate the ratio $G_3(\theta_s,\phi_s)/G_1$ by numerical
integration of the expression (\ref{snint}). The results for all
sky direction are shown in figure 6. We used the same Galactic model  as
 figure 3. The ratio becomes maximum value 1.56 around the Galactic
poles and the minimum 0.86 (not at the direction of the Galactic center).
The ratio $G_3/G_1$ is a convolution of the sky dependence of the response
function $F_{IJ}$ and the time   varying noise $S^B_{IJ}(t)$
whose profile is shown in figures 4 and 5. To interpret figure 6, let us
regard  the expression 
(\ref{snint}) as a integral of the function $S^B_{IJ}(t)^{-1}$ with the weight
$F_{IJ}$.   
The function $S^B_{IJ}(t)^{-1}$ becomes maximum when the $Z_D$-axis of the
detector is  nearly
oriented to the Galactic poles as shown in figures 3 and 4. At this
epoch the weight function $F_{IJ}$ is largest  to  the
direction of the Galactic poles and and smallest to the Galactic disk,
if we assume the matrix   $S^B_{IJ}(t)^{-1}$ in the form (3.6)
.
In this manner the map traces the Galactic structure well. We should also
notice that the overall  $\theta_s$ dependence of the ratio $G_3/G_1$ is
similar to the ratio
$G_2/G_1$, as easily expected.

We can make a similar argument for a short-lived signal such as the
ring down wave from a MBH binary. As mentioned above,  the best time   is
the autumn 
equinox  and  the best
source 
directions at that time   are  the Galactic poles (both $F_{IJ}$ and
$S^B_{IJ}(t)^{-1}$ are maximum). The worst season of the effective noise is
summer  when the orientation of the $Z_D$-axis is nearly on the Galactic
disk.  At this
configuration the response function $F_{IJ}$  takes the smallest
value to the Galactic poles that are now normal to the $Z_D$-axis. Thus
both of two elements $F_{IJ}$ and  
$(S^B_{IJ})^{-1}$ are nearly minimized. As a results the SNR (not
$SNR^2$)  at the direction of  Galactic poles changes by a factor of
4 in three month. A factor of  2 comes from the changes of the noise and
another factor of 2 comes from that of the response function. Both of  two
effects go to  the same direction to amplify the time modulation.

\section{summary}
In this paper we  have studied the anisotropies of the Galactic confusion
noise background and its effect on the data  analysis of LISA at the low
frequency regime  $f\lsim 3$mHz. In contrast to the traditional monopole
approximation the anisotropies induce the correlation of confusion
noises between the $A$ and $E$ modes of LISA, and the
effective noise level depends strongly on the time. In the followings we
briefly summarize our results.

The effective noise level of LISA becomes smallest around the autumn
equinox, 
and largest in summer for the configuration in figure 2. The difference
of the 
noise $\sqrt{S^B_{eff}}$ at these two epochs is a factor of 2.  The
detector plane of LISA at the autumn equinox is almost normal to  the
Galactic poles and  the effective noise level is
very close to  the global minimum  of the noise map given for the all
possible 
orientations of the detector.
The    time average of the effective noise is different from the
traditional  monopole approximation by less than $50\%$.

We have also analyzed the dependence of the SNR on the sky position of
sources with a fixed  distance and averaged orientation.
For a long lived source the dependence   after 1yr integration is weak,
and fluctuations  on the sky  are   $\sim 30\%$ level. This is 
mainly  
due to the angular average of the response function by annual rotation
of LISA.   
For a short lived source the SNR at a given direction changes strongly
with time. For example the SNR of a source  at the direction of the
Galactic pole
changes by a factor of $\sim4 $ in three months. A factor of  2 comes
from the 
modulation of the confusion noise level and another factor of  2 is from
that of  the angular response function of the detector.

\begin{acknowledgments}

The author  thanks Asantha Cooray and
Sterl Phinney  for valuable discussions.
This work is supported in part by NASA grant NAG5-10707.

\end{acknowledgments}

\appendix
\section{spherical harmonic expansion}
The noise matrix $S^B_{IJ}$ is determined by the orientation of the
coordinate  system $K_D ;(Z_D,Y_D,Z_D)$ (as defined in figure 1) that
would change with time. In this situation the spherical harmonic
expansion for a fixed coordinate system 
$K_0 ;(X_0,Y_0,Z_0)$ ({\it e.g.} the ecliptic coordinate in figure 2)
would be useful \cite{Cornish:2001hg}.  We can related two systems $K_D$
and $K_0$ by the Euler 
 angles $(\alpha,\beta,\gamma)$. Here $(\alpha,\beta)$ is the direction of
the $Z_D$-axis in the $K_0$ system, and $\gamma$ is essentially
corresponds to  the freedom of the $\phi_f$-rotation  in
eq.(\ref{trans}) around the $Z_D$-axis. We denote the noise matrix in 
the  $K_D$ system  as 
$S^B_{IJ} (\alpha,\beta,\gamma)$.
In the configuration of figure 2  we can parameterize the motion of LISA
with  $\alpha=2\pi T$,
$\beta=-\pi/3$ and $\gamma=-2\pi T+\gamma_0$ with some constant
$\gamma_0$ and the orbital time $T$.

The  angular dependence of the gravitational wave
intensity (eq.(\ref{gwaniso})) $B(\theta_0,\phi_0)$   is decomposed by the spherical harmonics
$Y_{lm}(\theta_0,\phi_0)$ as  
\beq
B(\theta_0,\phi_0)=\sum_{l,m} B_{lm} Y_{lm}(\theta_0,\phi_0).
\eeq
Here the coefficients  $B_{lm}$ is given as
\beq
B_{lm}=\lla lm|B \rra,
\eeq
where we  have used the traditional notation for the inner product
$
\lla a | b\rra \equiv \int d\Omega a^*(\theta_0,\phi_0)
b(\theta_0,\phi_0) ,
$ 
and
an abbreviation $|lm>\equiv |Y_{lm}(\theta_0,\phi_0)>$.
Our goal is to write down the matrix $S^B_{IJ}(\alpha,\beta,\gamma)$ with
using the coefficients $B_{lm}$.

First the matrix $S^B_{IJ}(\alpha,\beta,\gamma)$  is formally given as follows
\beq
S^B_{IJ}(\alpha,\beta,\gamma)=\lla F_{IJ}(\theta_0,\phi_0)|
U(\alpha,\beta,\gamma)^{-1}| B(\theta_0,\phi_0)\rra,
\eeq
where $U(\alpha,\beta,\gamma)$ is the rotation operator with the Eular
angles $\alpha,\beta$ and $\gamma$.
Using the identity $\sum_{lm}|lm><lm|=1$ for the complete sets $|lm>$, 
we have 
\beqa
S^B_{IJ}(\alpha,\beta,\gamma)&=&\sum_{lm}\sum_{l'm'}
\lla F_{IJ}(\theta_0,\phi_0)|lm\rra \lla lm
|U(\alpha,\beta,\gamma)^{-1}|l'm'\rra 
\lla l'm'|B(\theta_0,\phi_0)\rra,\\
&=&\sum_{lmm'}
\lla F_{IJ}(\theta_0,\phi_0)|lm\rra \lla lm
|U(\alpha,\beta,\gamma)^{-1}|lm'\rra 
\lla lm'|B(\theta_0,\phi_0)\rra. \label{formal}
\eeqa

Next we calculate the coefficients 
$
C_{IJlm}\equiv \lla F_{IJ}|lm\rra
$
for the angular dependence of the response functions. We have $C_{IJlm}=0$
for $l\ne 0,2$ and 4.  The followings are all of the non-vanishing 
 coefficients;
\beqa
C_{AA00}&=&C_{EE00}=\frac{2 {\sqrt \pi}}{5},\\
C_{AA20}&=&C_{EE20}=\frac47 {\sqrt \frac{\pi}5},\\
C_{AA40}&=&C_{EE40}=\frac{\sqrt \pi}{105},\\
C_{AA44}&=&C_{AA4-4}=-C_{EE44}=-C_{EE4-4}=\frac13 {\sqrt
\frac{\pi}{70}},\\
C_{AE44}&=&-C_{AE4-4}=\frac{i}3 {\sqrt \frac{\pi}{70}}.
\eeqa
Then eq.(\ref{formal}) is simplified to
\beqa
S^B_{IJ}(\alpha,\beta,\gamma)&=& B_{00} \lla
00|U(\alpha,\beta,\gamma)^{-1}|00\rra 
C_{IJ00 }
+\sum_m B_{2m} \lla 20|U(\alpha,\beta,\gamma)^{-1}|2m\rra
C_{IJ20 }  \nonumber \\
& &+\sum_m B_{4m}  \lla 40|U(\alpha,\beta,\gamma)^{-1}|4m\rra
C_{IJ40}+  \sum_m B_{4m} \lla 44|U(\alpha,\beta,\gamma)^{-1}|4m\rra
C_{IJ44} \\
& &+ \sum_m B_{4m} \lla 4-4|U(\alpha,\beta,\gamma)^{-1}|4m\rra
C_{IJ4-4}.
  \nonumber
\eeqa
For further  calculation we use the relation between the matrix elements 
$\lla lm|U(\alpha,\beta,\gamma)^{-1}|m'\rra$ and the spin-weight
spherical harmonics  ${}_sY_{lm} (\alpha,\beta)$ as
\beq
\lla lm|U(\alpha,\beta,\gamma)^{-1}| lm' \rra=e^{i\gamma
m} {}_{-m}
Y_{lm'} (\beta,\alpha) \sqrt{\frac{4\pi}{2l+1}}.
\eeq
After some calculation we finally obtain the explicit forms for the
matrix $S^B_{IJ}(\alpha,\beta,\gamma)$ as 
\beqa
S^B_{AA}&=&\frac{2 {\sqrt \pi}}{5}B_{00}+
\frac{8\pi}{35}B_{2m}Y_{2m}(\beta,\alpha) 
+\frac{2\pi}{315}B_{4m}Y_{4m}(\beta,\alpha)
+\frac{2\pi}{9{\sqrt {70}}}B_{4m}\lkk
{}_{-4}Y_{4m}(\beta,\alpha)e^{4i\gamma}
+ {}_{4}Y_{4m}(\beta,\alpha)e^{-4i\gamma} \rkk \\
S^B_{EE}&=&\frac{2 {\sqrt \pi}}{5}B_{00}+
\frac{8\pi}{35}B_{2m}Y_{2m}(\beta,\alpha) 
+\frac{2\pi}{315}B_{4m}Y_{4m}(\beta,\alpha)
-\frac{2\pi}{9{\sqrt {70}}}B_{4m}\lkk
{}_{-4}Y_{4m}(\beta,\alpha)e^{4i\gamma}
+ {}_{4}Y_{4m}(\beta,\alpha)e^{-4i\gamma} \rkk, \\
S^B_{AE}&=&\frac{2\pi i}{9{\sqrt {70}}}B_{4m}\lkk
{}_{-4}Y_{4m}(\beta,\alpha)e^{4i\gamma}
- {}_{4}Y_{4m}(\beta,\alpha)e^{-4i\gamma} \rkk,
\eeqa
where the summation with respect to the index $m$ is implicitly assumed.
These symmetric expressions would be quite useful to  study how we can
estimate the anisotropies $B_{lm}$ form the 
data $S^B_{IJ}$.  The correlation analysis $S^B_{AE}$ can be used to
extract the hexadecupole ($l=4$) mode and the freedom of adjusting $\gamma$
would be important here. Detailed studies on this issue would be presented
elsewhere.

We can easily confirm that both  ${\rm tr}S^B_{IJ}$ and $\det S^B_{IJ}$ do not
depend on the angle $\gamma$, as expected.  We can derive some expressions for the
eigen values $\lambda_1$ and  $\lambda_2$ of the noise matrix $S^B_{IJ}$. With
notation $(a(\beta,\alpha))_{av}$ for 
angular average of a function $a(\beta,\alpha)$  over a unit sphere we have
\beqa
\frac12(\lambda_1+\lambda_2)_{av}&=&\frac{2 {\sqrt \pi}}{5}B_{00},\\
\frac14((\lambda_1+\lambda_2)^2)_{av}&=&\frac{4\pi}{25} B_{00}^2
+\frac{16\pi}{35^2} \sum_m |B_{2m}|^2+\frac{\pi}{315^2} \sum_m
|B_{4m}|^2, \\ 
\frac14((\lambda_1- \lambda_2)^2)_{av}&=&
\frac{2\pi}{2835} \sum_m
|B_{4m}|^2.
\eeqa

\begin{figure}
  \begin{center}
\epsfxsize=10.cm
\begin{minipage}{\epsfxsize} \epsffile{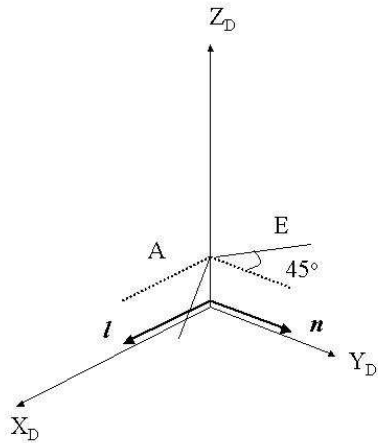} \end{minipage}
 \end{center}
  \caption{ Definition of the detector coordinate $(X_D,Y_D,Z_D)$. The
 $Z_D$-axis is normal  to the detector plane. A and E modes can be
 regarded as two L-shaped detectors rotated by $45^\circ$.}
\end{figure}

\begin{figure}
  \begin{center}
\epsfxsize=15.cm
\begin{minipage}{\epsfxsize} \epsffile{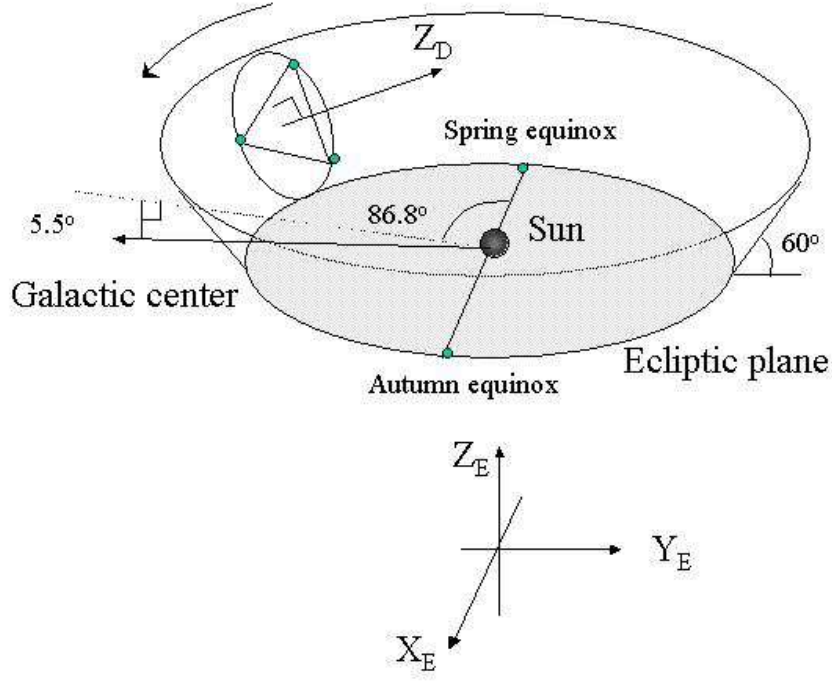} \end{minipage}
 \end{center}
  \caption{Configuration of LISA and the definition of the ecliptic
 coordinate $(X_E, Y_E, Z_E)$. The $Z_E$-axis is normal to the ecliptic
 plane and  $X_E$-axis is oriented to   the autumn
 equinox. The direction of the Galactic center is $\theta=95.5^\circ$ and $\phi=-93.2^\circ$
 in this coordinate. With this configuration
 the azimuthal angle for  the orientation of the$Z_D$-axis and  that for
 the direction of 
 the 
 detector (from the Sun) differ by  $180^\circ$.} 
\end{figure}

\begin{figure}
  \begin{center}
\epsfxsize=15.cm
\begin{minipage}{\epsfxsize} \epsffile{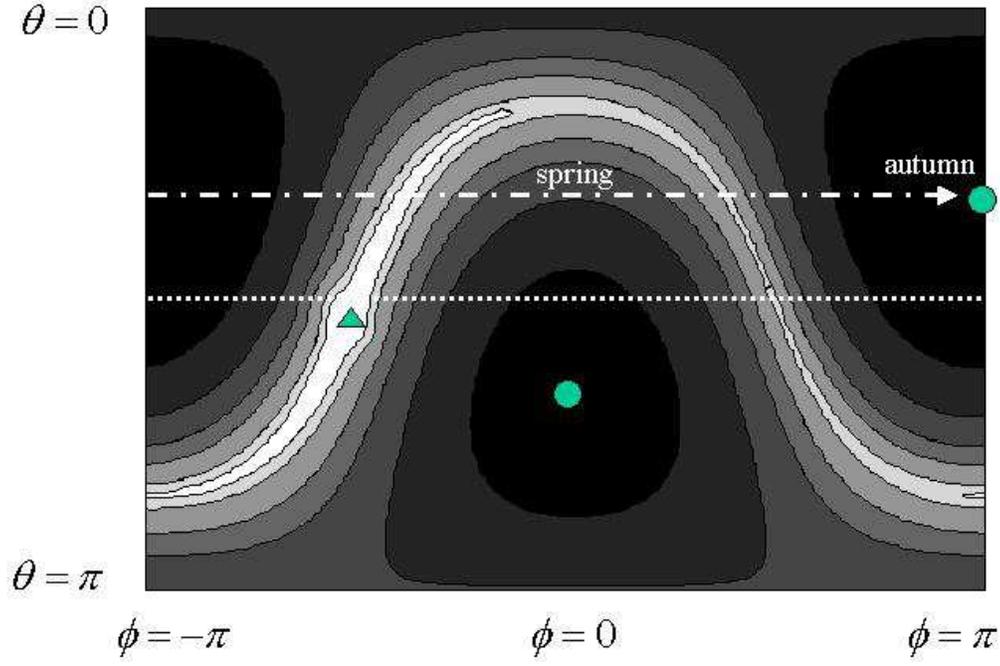} \end{minipage}
 \end{center}
  \caption{ The gravitational wave luminosity $B(\theta,\phi)$ on the sky
  with the ecliptic coordinate. The BGS model  is used
 for the Galactic binary distribution.  The direction of the
 Galactic center $(\theta_{GC},\phi_{GC})$ is given by the triangle, and the Galactic poles by
 circles.  The contours correspond to
 $\log_{10}(B(\theta,\phi )/B(\theta_{GC},\phi_{GC}))=-2$ (white) to $-3$
 (black)   
 with  interval 0.2.  The $Z_D$-axis of LISA moves along the dash-dotted line.}
\end{figure}

\begin{figure}
  \begin{center}
\epsfxsize=15.cm
\begin{minipage}{\epsfxsize} \epsffile{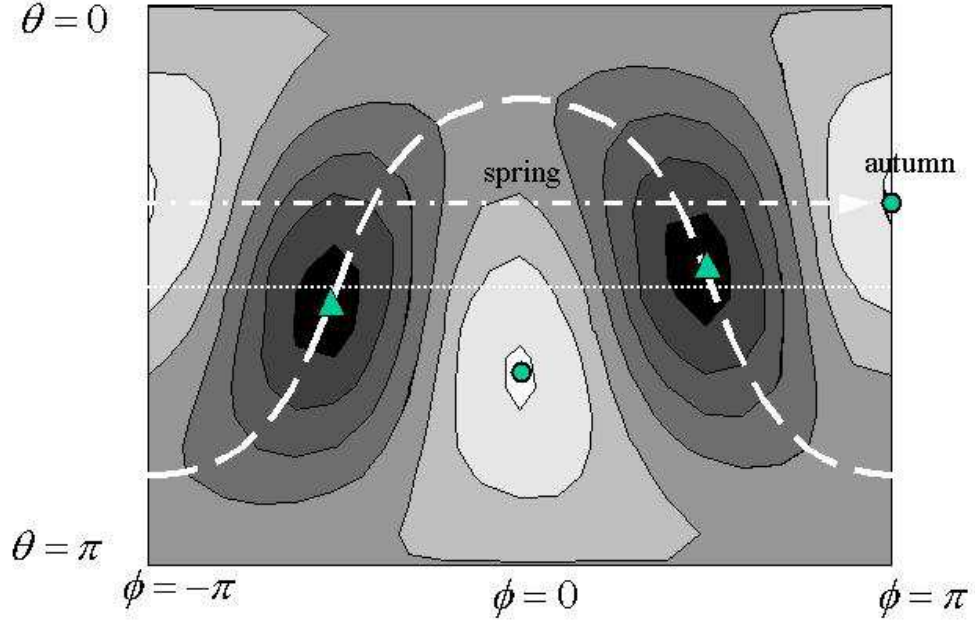} \end{minipage}
 \end{center}
  \caption{All sky effective noise map in the ecliptic coordinate. The
 effective noise 
 $S^B_{eff}/S^B_{iso}$ is plotted as a function of the orientation
 $(\theta,\phi)$ of the $Z_D$-axis of the detector.  Contours correspond
 to 
 $S^B_{eff}/S^B_{iso}=0.4$ (white) to 1.6 (black) with 0.2 interval.
 The dotted line is the 
 ecliptic plane. The $Z_D$-axis of LISA moves along   the dash-dotted
 line.  The
 direction of the  
 Galactic center and Galactic poles are  given by the triangles and
 circles. The Galactic disk plane is given by the long dashed line.}
\end{figure}

\begin{figure}
  \begin{center}
\epsfxsize=8.cm
\begin{minipage}{\epsfxsize} \epsffile{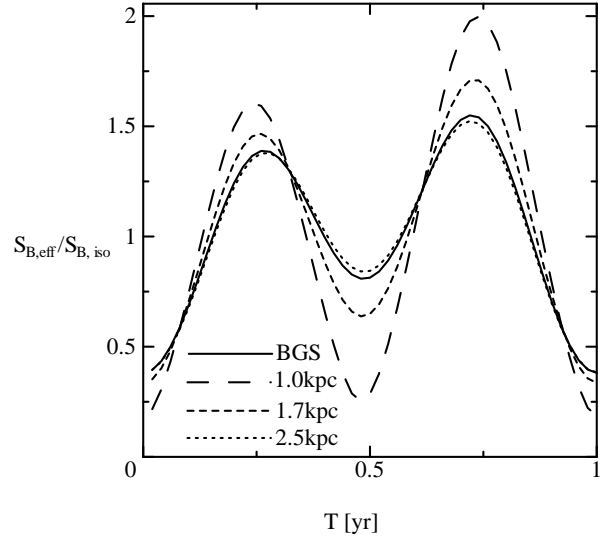} \end{minipage}
 \end{center}
  \caption{ Annual modulation of the effective noise
 $S^B_{eff}/S^B_{iso}$ as a function of the orbital phase $T$ of
 LISA. The solid line is for the BGS model. Other three 
 curves correspond to  the different scale length $R_s$ of the Galactic distribution (\ref{disk}). }
\end{figure}

\begin{figure}
  \begin{center}
\epsfxsize=15.cm
\begin{minipage}{\epsfxsize} \epsffile{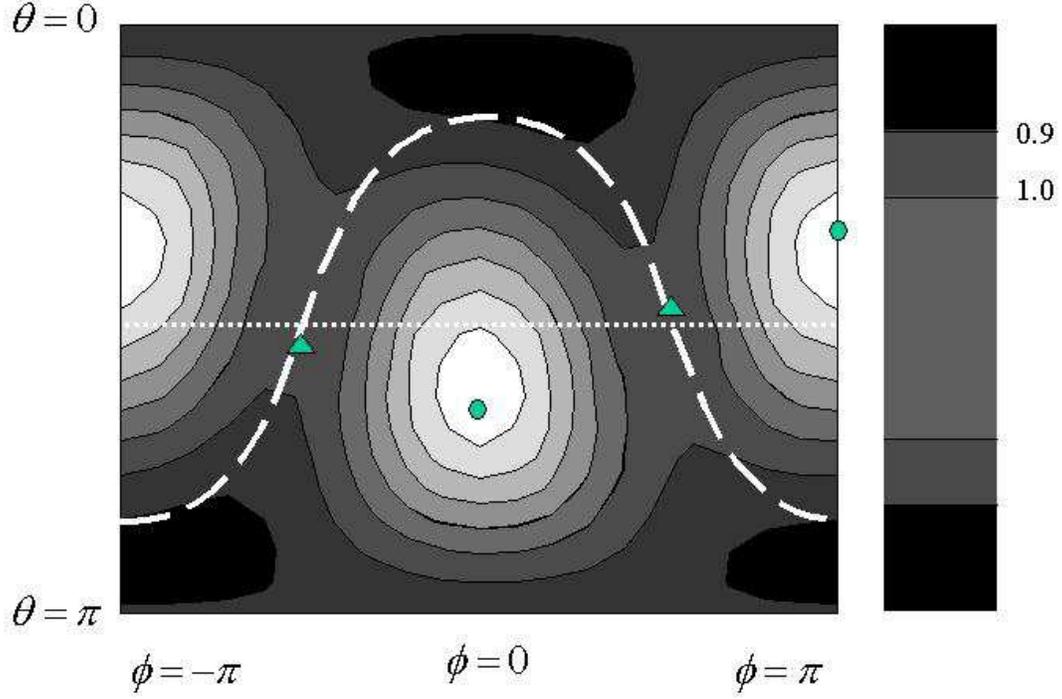} \end{minipage}
 \end{center}
  \caption{Directional  dependence of $SNR^2$ in the ecliptic
 coordinate after 1yr integration. The left panel is the ratio 
 $G_3(\theta,\phi)/G_1$. The right small panel is $G_2(\theta)/G_1$
 that does not depend on the azimuthal angle $\phi$ as eq.(4.2). Contours
 correspond to 
 $S^B_{eff}/S^B_{iso}=0.9$ (black) to 1.5 (white) with 0.1 interval. The
 directions of the Galactic poles are given by the circles.  The
 Galactic disk plane is given by the long dashed line. By definition
 this line is $90^\circ$ apart from the poles. This map
 does not depend on the possible two  configurations of LISA.}
\end{figure}

\end{document}